
%
%
%
%
\def\unredoffs{\hoffset-.14truein\voffset-.2truein} 
 
%
%
\newbox\leftpage \newdimen\fullhsize \newdimen\hstitle \newdimen\hsbody
\tolerance=1000\hfuzz=2pt
\catcode`\@=11 
%
\magnification=1095\unredoffs\baselineskip=16pt plus 2pt minus 1pt
\hsbody=\hsize \hstitle=\hsize 
%
%
%
\newcount\yearltd\yearltd=\year\advance\yearltd by -1900

%
%

\def\draftmode{\message{ DRAFTMODE }\def\draftdate{{\rm preliminary draft:
\number\month/\number\day/\number\yearltd\ \ \hourmin}}%
\headline={\hfil\draftdate}\writelabels\baselineskip=16pt plus 2pt minus 2pt
 {\count255=\time\divide\count255 by 60 \xdef\hourmin{\number\count255}
  \multiply\count255 by-60\advance\count255 by\time
  \xdef\hourmin{\hourmin:\ifnum\count255<10 0\fi\the\count255}}}
\def\nolabels{\def\wrlabeL##1{}\def\eqlabeL##1{}\def\reflabeL##1{}}
\def\writelabels{\def\wrlabeL##1{\leavevmode\vadjust{\rlap{\smash%
{\line{{\escapechar=` \hfill\rlap{\sevenrm\hskip.03in\string##1}}}}}}}%
\def\eqlabeL##1{{\escapechar-1\rlap{\sevenrm\hskip.05in\string##1}}}%
\def\reflabeL##1{\noexpand\llap{\noexpand\sevenrm\string\string\string##1}}}
\nolabels
%
\global\newcount\secno \global\secno=0
\global\newcount\meqno \global\meqno=1
\def\newsec#1{\global\advance\secno by1\message{(\the\secno. #1)}
\global\subsecno=0\eqnres@t\noindent{\bf\the\secno. #1}
\writetoca{{\secsym} {#1}}\par\nobreak\medskip\nobreak}
\def\eqnres@t{\xdef\secsym{\the\secno.}\global\meqno=1\bigbreak\bigskip}
\def\sequentialequations{\def\eqnres@t{\bigbreak}}\xdef\secsym{}
\global\newcount\subsecno \global\subsecno=0
\def\subsec#1{\global\advance\subsecno by1\message{(\secsym\the\subsecno. #1)}
\ifnum\lastpenalty>9000\else\bigbreak\fi
\noindent{\bf\secsym\the\subsecno. #1}\writetoca{\string\quad 
{\secsym\the\subsecno.} {#1}}\par\nobreak\medskip\nobreak}
\def\appendix#1#2{\global\meqno=1\global\subsecno=0\xdef\secsym{\hbox{#1.}}
\bigbreak\bigskip\noindent{\bf Appendix #1. #2}\message{(#1. #2)}
\writetoca{Appendix {#1.} {#2}}\par\nobreak\medskip\nobreak}
%
%
\def\eqnn#1{\xdef #1{(\secsym\the\meqno)}\writedef{#1\leftbracket#1}%
\global\advance\meqno by1\wrlabeL#1}
\def\eqna#1{\xdef #1##1{\hbox{$(\secsym\the\meqno##1)$}}
\writedef{#1\numbersign1\leftbracket#1{\numbersign1}}%
\global\advance\meqno by1\wrlabeL{#1$\{\}$}}
\def\eqn#1#2{\xdef #1{(\secsym\the\meqno)}\writedef{#1\leftbracket#1}%
\global\advance\meqno by1$$#2\eqno#1\eqlabeL#1$$}
%
\newskip\footskip\footskip14pt plus 1pt minus 1pt 
\def\footnotefont{\ninepoint}\def\f@t#1{\footnotefont #1\@foot}
\def\f@@t{\baselineskip\footskip\bgroup\footnotefont\aftergroup\@foot\let\next}
\setbox\strutbox=\hbox{\vrule height9.5pt depth4.5pt width0pt}
\global\newcount\ftno \global\ftno=0
\def\foot{\global\advance\ftno by1\footnote{$^{\the\ftno}$}}
%
\newwrite\ftfile   
\def\footend{\def\foot{\global\advance\ftno by1\chardef\wfile=\ftfile
$^{\the\ftno}$\ifnum\ftno=1\immediate\openout\ftfile=foots.tmp\fi%
\immediate\write\ftfile{\noexpand\smallskip%
\noexpand\item{f\the\ftno:\ }\pctsign}\findarg}%
\def\footatend{\vfill\eject\immediate\closeout\ftfile{\parindent=20pt
\centerline{\bf Footnotes}\nobreak\bigskip\input foots.tmp }}}
\def\footatend{}
%
%
\global\newcount\refno \global\refno=1
\newwrite\rfile
\def\ref{[\the\refno]\nref}
\def\nref#1{\xdef#1{[\the\refno]}\writedef{#1\leftbracket#1}%
\ifnum\refno=1\immediate\openout\rfile=refs.tmp\fi
\global\advance\refno by1\chardef\wfile=\rfile\immediate
\write\rfile{\noexpand\item{#1\ }\reflabeL{#1\hskip.31in}\pctsign}\findarg}
\def\findarg#1#{\begingroup\obeylines\newlinechar=`\^^M\pass@rg}
{\obeylines\gdef\pass@rg#1{\writ@line\relax #1^^M\hbox{}^^M}%
\gdef\writ@line#1^^M{\expandafter\toks0\expandafter{\striprel@x #1}%
\edef\next{\the\toks0}\ifx\next\em@rk\let\next=\endgroup\else\ifx\next\empty%
\else\immediate\write\wfile{\the\toks0}\fi\let\next=\writ@line\fi\next\relax}}
\def\striprel@x#1{} \def\em@rk{\hbox{}} 
\def\lref{\begingroup\obeylines\lr@f}
\def\lr@f#1#2{\gdef#1{\ref#1{#2}}\endgroup\unskip}
\def\semi{;\hfil\break}
\def\addref#1{\immediate\write\rfile{\noexpand\item{}#1}} 
\def\startrefs#1{\immediate\openout\rfile=refs.tmp\refno=#1}
\def\xref{\expandafter\xr@f}\def\xr@f[#1]{#1}
\def\refs#1{\count255=1[\r@fs #1{\hbox{}}]}
\def\r@fs#1{\ifx\und@fined#1\message{reflabel \string#1 is undefined.}%
\nref#1{need to supply reference \string#1.}\fi%
\vphantom{\hphantom{#1}}\edef\next{#1}\ifx\next\em@rk\def\next{}%
\else\ifx\next#1\ifodd\count255\relax\xref#1\count255=0\fi%
\else#1\count255=1\fi\let\next=\r@fs\fi\next}
%

%
\newwrite\ffile\global\newcount\figno \global\figno=1
\def\fig{fig.~\the\figno\nfig}
\def\nfig#1{\xdef#1{fig.~\the\figno}%
\writedef{#1\leftbracket fig.\noexpand~\the\figno}%
\ifnum\figno=1\immediate\openout\ffile=figs.tmp\fi\chardef\wfile=\ffile%
\immediate\write\ffile{\noexpand\medskip\noexpand\item{Fig.\ \the\figno. }
\reflabeL{#1\hskip.55in}\pctsign}\global\advance\figno by1\findarg}
\def\xfig{\expandafter\xf@g}\def\xf@g fig.\penalty\@M\ {}
\def\figs#1{figs.~\f@gs #1{\hbox{}}}
\def\f@gs#1{\edef\next{#1}\ifx\next\em@rk\def\next{}\else
\ifx\next#1\xfig #1\else#1\fi\let\next=\f@gs\fi\next}
\newwrite\lfile
{\escapechar-1\xdef\pctsign{\string\%}\xdef\leftbracket{\string\{}
\xdef\rightbracket{\string\}}\xdef\numbersign{\string\#}}

\def\writestop{\def\writestoppt{\immediate\write\lfile{\string\pageno%
\the\pageno\string\startrefs\leftbracket\the\refno\rightbracket%
\string\def\string\secsym\leftbracket\secsym\rightbracket%
\string\secno\the\secno\string\meqno\the\meqno}\immediate\closeout\lfile}}
\def\writestoppt{}\def\writedef#1{}
\def\seclab#1{\xdef #1{\the\secno}\writedef{#1\leftbracket#1}\wrlabeL{#1=#1}}
\def\subseclab#1{\xdef #1{\secsym\the\subsecno}%
\writedef{#1\leftbracket#1}\wrlabeL{#1=#1}}
\newwrite\tfile \def\writetoca#1{}
\def\leaderfill{\leaders\hbox to 1em{\hss.\hss}\hfill}
\def\writetoc{\immediate\openout\tfile=toc.tmp 
   \def\writetoca##1{{\edef\next{\write\tfile{\noindent ##1 
   \string\leaderfill {\noexpand\number\pageno} \par}}\next}}}

%
\catcode`\@=12 
%
\edef\tfontsize{\ifx\answ\bigans scaled\magstep3\else scaled\magstep4\fi}
 \tfontsize  \tfontsize
 \tfontsize \font\titlei=cmmi10 \tfontsize
\font\titleis=cmmi7 \tfontsize \font\titleiss=cmmi5 \tfontsize
\font\titlesy=cmsy10 \tfontsize \font\titlesys=cmsy7 \tfontsize
\font\titlesyss=cmsy5 \tfontsize  \tfontsize
\skewchar\titlei='177 \skewchar\titleis='177 \skewchar\titleiss='177
\skewchar\titlesy='60 \skewchar\titlesys='60 \skewchar\titlesyss='60
 \ifx\answ\bigans\else scaled\magstep1\fi
\ifx\answ\bigans\else

 \font\absi=cmmi10 scaled\magstep1
\font\absis=cmmi7 scaled\magstep1 \font\absiss=cmmi5 scaled\magstep1
\font\abssy=cmsy10 scaled\magstep1 \font\abssys=cmsy7 scaled\magstep1
\font\abssyss=cmsy5 scaled\magstep1 
\skewchar\absi='177 \skewchar\absis='177 \skewchar\absiss='177
\skewchar\abssy='60 \skewchar\abssys='60 \skewchar\abssyss='60
\fi
\font\ninerm=cmr9 \font\sixrm=cmr6 \font\ninei=cmmi9 \font\sixi=cmmi6 
\font\ninesy=cmsy9 \font\sixsy=cmsy6 \font\ninebf=cmbx9 
\font\nineit=cmti9 \font\ninesl=cmsl9 \skewchar\ninei='177
\skewchar\sixi='177 \skewchar\ninesy='60 \skewchar\sixsy='60 
\def\ninepoint{\def\rm{\fam0\ninerm}
\textfont0=\ninerm \scriptfont0=\sixrm \scriptscriptfont0=\fiverm
\textfont1=\ninei \scriptfont1=\sixi \scriptscriptfont1=\fivei
\textfont2=\ninesy \scriptfont2=\sixsy \scriptscriptfont2=\fivesy
\textfont\itfam=\ninei \def\it{\fam\itfam\nineit}\def\sl{\fam\slfam\ninesl}%
\textfont\bffam=\ninebf \def\bf{\fam\bffam\ninebf}\rm} 
%
%

\hyphenation{anom-aly anom-alies coun-ter-term coun-ter-terms}
\def\inv{^{\raise.15ex\hbox{${\scriptscriptstyle -}$}\kern-.05em 1}}

\def\Dsl{\,\raise.15ex\hbox{/}\mkern-13.5mu D} 
\def\dsl{\raise.15ex\hbox{/}\kern-.57em\partial}

\def\lspace{\ifx\answ\bigans{}\else\qquad\fi}
\def\lbspace{\ifx\answ\bigans{}\else\hskip-.2in\fi} 
\def\boxeqn#1{\vcenter{\vbox{\hrule\hbox{\vrule\kern3pt\vbox{\kern3pt
	\hbox{${\displaystyle #1}$}\kern3pt}\kern3pt\vrule}\hrule}}}
\def\mbox#1#2{\vcenter{\hrule \hbox{\vrule height#2in
		\kern#1in \vrule} \hrule}}  
%

\def\darr#1{\raise1.5ex\hbox{$\leftrightarrow$}\mkern-16.5mu #1}

\def\half{{\textstyle{1\over2}}} 
\def\roughly#1{\raise.3ex\hbox{$#1$\kern-.75em\lower1ex\hbox{$\sim$}}}

\def\p2inf{\mathrel{\mathop{\sim}\limits_{\scriptscriptstyle
{p^2 \rightarrow \infty }}}}
\def\kap2inf{\mathrel{\mathop{\sim}\limits_{\scriptscriptstyle
{\kappa \rightarrow \infty }}}}
\def\x2inf{\mathrel{\mathop{\sim}\limits_{\scriptscriptstyle
{x \rightarrow \infty }}}}
\def\Lam2inf{\mathrel{\mathop{\sim}\limits_{\scriptscriptstyle
{\Lambda \rightarrow \infty }}}}
\def\frac#1#2{{{#1}\over {#2}}}
\def\half{\hbox{${1\over 2}$}}

\def\Gev{{\rm GeV}}

\def\lsim{\mathrel{mathpalette\@v1000ersim<}}
\def\gsim{\mathrel{mathpalette\@versim>}}

\catcode`@=11 
\def\slash#1{\mathord{\mathpalette\c@ncel#1}}
 \def\c@ncel#1#2{\ooalign{$\hfil#1\mkern1mu/\hfil$\crcr$#1#2$}}
\def\lsim{\mathrel{\mathpalette\@versim<}}
\def\gsim{\mathrel{\mathpalette\@versim>}}
 \def\@versim#1#2{\lower0.2ex\vbox{\baselineskip\z@skip\lineskip\z@skip
       \lineskiplimit\z@\ialign{$\m@th#1\hfil##$\crcr#2\crcr\sim\crcr}}}
\catcode`@=12 

\def\PR{{\it Phys.~Rev.~}}

\def\NP{{\it Nucl.~Phys.~}}
\def\PL{{\it Phys.~Lett.~}}
\def\PRep{{\it Phys.~Rep.~}}

\def\SJNP{{\it Sov.~Jour.~Nucl.~Phys.~}}
\def\ZP{{\it Zeit.~Phys.~}}

\def\vol#1{{\bf #1}}
\def\vyp#1#2#3{\vol{#1} (#2) #3}

\def\Asl{\raise.15ex\hbox{/}\mkern-11.5mu A}
\def\psl{\lower.12ex\hbox{/}\mkern-9.5mu p}
\def\qsl{\lower.12ex\hbox{/}\mkern-9.5mu q}
\def\rsl{\lower.03ex\hbox{/}\mkern-9.5mu r}
\def\ksl{\raise.06ex\hbox{/}\mkern-9.5mu k}


\pageno=0\nopagenumbers\tolerance=10000\hfuzz=5pt
\line{\hfill OUTP-99??P}
\vskip 36pt
\centerline{\bf Explicit Calculation Of The Running Coupling BFKL}
\vskip 6pt
\centerline{\bf Anomalous Dimension.}
\vskip 36pt
\centerline{Robert~S.~Thorne}
\vskip 12pt
\centerline{\it Jesus College and Theoretical Physics,}
\centerline{\it University of Oxford, Oxford, Oxon., OX1 3DW, U.K.}
\vskip 0.9in
{\narrower\baselineskip 10pt
\centerline{\bf Abstract}
\medskip
I calculate the anomalous dimension governing the $Q^2$ evolution 
of the gluon (and structure functions) coming from the running coupling
BFKL equation. This may be expressed in an exact analytic form, up to
a small ultraviolet renormalon contribution, and hence the
corresponding splitting function may be determined precisely. 
Rather surprisingly it is most efficient to expand the gluon
distribution in powers of $\alpha_s(Q^2)$ rather than use the traditional
expansion where all orders of $\alpha_s\ln(1/x)$ are kept on an equal
footing. The
anomalous dimension is very different from that obtained from the fixed
coupling equation, and leads to a powerlike behaviour for the
splitting function as $x \to
0$ which is far weaker, i.e. $\sim x^{-0.2}$. The NLO corrections to
the anomalous dimension are rather small, unlike the fixed
coupling case, and a stable perturbative expansion is obtained.}
   
\vskip 0.7in
\line{\hfill}
\line{December 1999\hfill}
\vfill\eject
\footline={\hss\tenrm\folio\hss}



\newsec{Introduction.}

Small $x$ physics has been an active area of theoretical research in
the past few years, largely due to the first data for $x<0.005$ being
obtained by the HERA experiments \ref\hone{H1 
collaboration: S. Aid {\it et al.},\NP  \vyp{B470}{1996}{3}\semi
H1 collaboration: C. Adloff {\it et al.}, \NP
\vyp{B497}{1997}{3}\semi 
H1 collaboration: C. Adloff {\it et al.}, {\tt hep-ex/9908059}.}
\ref\zeus{ZEUS collaboration: M. Derrick {\it et al},
\ZP  \vyp{C69}{1996}{607}\semi ZEUS collaboration: M. Derrick {\it et al},
\ZP  \vyp{C72}{1996}{399}.}. The crux of the debate has been
whether the standard DGLAP approach based on renormalization group
equations and conventionally ordered simply in powers of
$\alpha_s(Q^2)$, or the BFKL equation \ref\BFKL{L.N. Lipatov, 
\SJNP \vyp{23}{1976}{338}\semi
E.A. Kuraev, L.N. Lipatov and V.S. Fadin, {\it Sov.~Jour. JETP} 
\vyp{45}{1977}{199}\semi
Ya. Balitskii and L.N. Lipatov, \SJNP \vyp{28}{1978}{6}.}, 
which sums leading logarithms
in $(1/x)$, is most effective, and/or whether the two approaches need
to be combined in some way. While the DGLAP approach has been
relatively successful, albeit with some significant problems
(a valence-like input gluon, undershooting for $x \sim 0.01$ at the highest
$Q^2$, see \ref\mrst{A.D. Martin {\it et al}, {\it Eur. Jour. Phys.},
{\bf C4} (1998) 463.}), the
original BFKL prediction of a behaviour of the form $x^{-\lambda}$ at
small $x$, with $\lambda \sim 0.5$, was clearly ruled out. A
combination of the two approaches, using the BFKL equation to
supplement the Altarelli-Parisi splitting functions with higher terms
of the form $\alpha_s^{n+1} \ln^n(1/x)$ had some success (so long as
one avoided factorization scheme ambiguities by working in physical quantities)
\ref\LORSC{R.S. Thorne, \PL \vyp{B392}{1997}{463}; \NP 
\vyp{B512}{1998}{323}.}, but this was
difficult to sustain with the most recent data. Moreover, the subject
was thrown into some confusion by the calculation of the NLO
correction to the BFKL equation \ref\NLOBFKLlf{V.S. Fadin and
L.N. Lipatov, \PL \vyp{B429}{1998}{127}, and
references therein.}\ref\NLOBFKLcc{G. Camici and M. Ciafaloni, 
\PL \vyp{B430}{1998}{349}, and references therein.}.  

In order to illustrate
this I begin with a brief discussion of the LO fixed coupling 
BFKL equation. Working in moment
space, i.e. defining the moment of the structure function by
\eqn\melltranssf{{\cal F}(N,Q^2)= \int_0^1\,x^{N-1} F(x,Q^2)dx,}
and similarly for the parton distributions (scaled by $x$), the BFKL
equation is 
\eqn\bfkli{f(k^2, \bar\alpha_s/N)=f_I(k^2, Q_0^2)+{\bar
\alpha_s \over N}
\int_{0}^{\infty}{dq^2 \over q^2}K_{0}(q^2,k^2)f(q^2),}
where $f(k^2, \bar \alpha_s/N)$ is the unintegrated gluon four-point
function, $f_I(k^2,Q_0^2)$ is the zeroth order input, $\bar \alpha_s =(3/\pi)
\alpha_s$, and the LO kernel is defined by 
\eqn\kzero{K_0(q^2,k^2)f(q^2)= k^2
\biggl( {f(q^2)-f(k^2) \over \mid k^2-q^2\mid}
+{f(k^2) \over (4q^4+k^4)^{\half}}\biggr).}
It is convenient to define the input by $f_I(k^2,Q_0^2) = \delta(k^2-Q_0^2)$,
where in the case of deep-inelastic scattering, where one end of the
gluon ladder is at a hard scale $Q^2$, while the other end is formally
on-shell, $Q_0^2$ is just a collinear regularization which we let $\to
0$ ultimately. The ``gluon structure function'' is then given by          
\eqn\gluondef{{\cal G}(Q^2,N)=\int_{0}^{Q^2}{dk^2\over k^2} f(N,k^2,Q_0^2)
\times g_B(N,Q_0^2),}
where $g_B(N,Q_0^2)$ is a bare, nonperturbative gluon density in the
proton which implicitly absorbs the collinear divergences in $f(k^2)$. 
The BFKL equation is most easily solved by taking the Mellin
transformation to $\gamma$-space, i.e. 
\eqn\mellin{\tilde f(\gamma,N)=
\int_{0}^{\infty}d k^2 (k^2)^{-1-\gamma} f(k^2, N),}
where it reduces to 
\eqn\bfklii{\tilde f(\gamma,N)=\tilde f_I(\gamma, Q_0^2)+
(\bar \alpha_s/N)  \chi_0(\gamma)
\tilde f(\gamma, N), }
where $\tilde f(\gamma, Q_0^2)=\exp(-\gamma \ln (Q_0^2))$
and $\chi(\gamma)$ is the characteristic function
\eqn\kergam{\chi_0(\gamma)=2\psi(1)-\psi(\gamma)-\psi(1-\gamma).}
A little manipulation leads to the expression
\eqn\invmell{{\cal G}(Q^2,N)={1\over 2\pi i}
\int_{\half-i\infty}^{\half+i\infty}
d\gamma \exp(\gamma \ln(Q^2/Q_0^2)) {g_B(N,Q_0^2) \over \gamma
(1-(\bar\alpha_s/N) \chi_0(\gamma))}.}
This inverse transformation is dominated by the leading pole at
$1-(\bar\alpha_s/N)
\chi_0(\gamma)=0$, and the solution is
\eqn\soliv{{\cal G}(Q^2,N)= {1 \over -(\bar\alpha_s/N)\gamma_0
\chi'_0(\gamma_0)}\biggl({Q^2\over
Q_0^2}\biggr)^{\gamma_0}g_B(N,Q_0^2).}
The anomalous dimension $\gamma_0(\bar\alpha_s/N)$ may be transformed
to $x$-space as a power series in $\bar \alpha_s\ln(1/x)$, and has a
branch point at $N=4\ln 2 \bar\alpha_s$ (at which $\gamma\to \half$) 
leading to asymptotic small $x$ behaviour for the splitting function
\eqn\split{xP^0(x) \to 0.07\bar \alpha_sx^{-\lambda}/(\bar\alpha_s 
\ln(1/x))^{3/2}.}

One can, if one ignores the running of the coupling, proceed through
exactly the same sort of arguments including the NLO correction to the
kernel. In this case the ``intercept'' for the splitting function is
shifted from  $\lambda=4\ln 2\bar\alpha_s$ to $\lambda=4\ln
2\bar\alpha_s(1-6.5\bar\alpha_s)$. This is clearly a huge correction,
and implies the breakdown of the perturbative expansion for this
quantity. Moreover, the power series for the splitting function is
dominated by the NLO corrections at all values of $x$ below about
$x=0.01$. For example, using the formulae in \NLOBFKLlf\
the first few terms in the power series for $P(x)$ go like
\eqn\unstab{\eqalign{xP(x)=&\bar\alpha_s+2.4\bar\alpha_s^4\xi^3/6+
2.1\bar\alpha_s^6\xi^5/120+ \cdots\cr
&-\bar\alpha_s(0.43\bar\alpha_s
+1.6\bar\alpha_s^2\xi+11.7\bar\alpha_s^3
\xi^2/2+13.3\bar\alpha_s^4\xi^3/6+39.7\bar\alpha_s^5\xi^4/24
+169.4\bar\alpha_s^6\xi^5/120+\cdots),\cr}}
where $\xi=\ln(1/x)$.
Clearly, the size of the coefficients more than compensates for the extra
power of $\alpha_s$. 
More careful calculations (discussed a little more below), 
including the running coupling
but using saddle-point approximations imply the same result, but with
the coupling being evaluated at $Q^2$. This does nothing to aid the 
convergence, particularly at values of $Q^2\sim 1\Gev^2$ where the
perturbative analysis of structure function evolution does take place.
      
Hence, this NLO correction left the whole question of how to 
address the evolution of
structure functions at small $x$ in question, and various alternatives
have been proposed to help stabilize the calculation.
In this paper I demonstrate that the correct way in which to calculate
the $Q^2$ evolution at small $x$ is indeed to supplement the conventional AP
splitting functions with higher order corrections obtained from the
solution to the BFKL equation, but to take the running of the coupling
constant fully into account in this equation. The details of this are
presented below.

\newsec{BFKL Equation for Running Coupling.}

Beyond leading order it is impossible to ignore the running of the
coupling, since at NLO ultraviolet
regularization is required, resulting in a $\ln(k^2/\mu_R^2)$ term
where $\mu_R$ is the renormalization scale. Such a term may be eliminated
by using the running coupling constant evaluated at the scale
$k^2$. Since this is unavoidably forced upon us at NLO, it seems
sensible to consider the fixed coupling LO BFKL equation as just a
model which would apply in a conformally invariant world, and more
realistically to work with the BFKL equation with running coupling
\ref\glr{L.V. Gribov, E.M. Levin and 
M.G. Ryskin, \PRep \vyp{100}{1983}{1}.} 
\ref\liprun{L.N. Lipatov, {\it Sov. Phys. JETP} \vyp{63}{1986}{904}.}%
\nref\jan{J. Kwiecinski, \ZP \vyp{C29}{1985}{561}.}%
\nref\janpcoll{J.C. Collins
and J. Kwiecinski, \NP \vyp{B316}{1989}{307}.}
from the beginning. Doing this we obtain 
\eqn\bfklruni{f(k^2, Q_0^2, \bar\alpha_s(k^2)/N)=f_I(k^2, Q_0^2)+
{\bar \alpha_s(k^2) \over N}
\int_{0}^{\infty}{dq^2 \over q^2}K_{0}(q^2,k^2)f(q^2),}
where
\eqn\coupdef{\alpha_s=1/(\beta_0\ln(k^2/\Lambda^2)),}
$\beta_0 = (11-2N_f/3)/(4\pi)$, and $N_f$ is the number of active 
flavours. 
One can solve this equation in the 
same type of way as for the fixed coupling case, i.e. take the Mellin 
transformation with respect to $(k^2/\Lambda^2)$. 
It is most convenient first to multiply through by 
$\ln(k^2/\Lambda^2)$, and then obtain 
\eqn\bfklrunii{{d\tilde f(\gamma,N)\over d \gamma}={d \tilde 
f_I(\gamma, Q_0^2) \over
d\gamma}-{1\over \bar\beta_0 N} \chi(\gamma)
\tilde f(\gamma, N),}
where $\bar \beta_0=(\pi\beta_0/3)$. 
Hence, the inclusion of the running coupling has completely changed
the form of 
our double Mellin space equation, turning it into a 
first order differential equation. This has a profound effect on the
form of the solutions. 
The equation may easily be solved 
giving,
\eqn\solruni{\tilde f(\gamma,N)=\exp(-X_0(\gamma,N)/(\bar\beta_0 N))
\int_{\gamma}^{\infty}
{d \tilde f_I(\tilde \gamma,N,Q_0^2)
\over d\tilde \gamma}\exp(X_0(\tilde \gamma)/(\bar\beta_0 N))d\tilde\gamma,}
where
\eqn\solrunii{X_0(\gamma)=
\int_{\half}^{\gamma}\chi_0(\hat\gamma)d\hat\gamma
\equiv \biggl(2\psi(1)(\gamma-\half)-
\ln\biggl({\Gamma(\gamma) \over \Gamma(1-\gamma)}\biggr)\biggr).}
The leading singularity in the $\gamma$ plane for
$\exp(-X_0(\gamma)/(\bar\beta_0 N))$, is 
cancelled by an integral from $0 \to \gamma$ of the integrand depending
on $\tilde \gamma$ \janpcoll, and so up to ${\cal O}(\Lambda^2/k^2)$ 
corrections \solruni\ simplifies to 
\eqn\solruniii{\tilde f(\gamma,N)=\exp(-X_0(\gamma)/
(\bar\beta_0 N))\int_{0}^{\infty}
{d\tilde f_I(\tilde\gamma,N,Q_0^2)
\over d\tilde \gamma}\exp(X_0(\tilde \gamma)/(\bar\beta_0 N))d\tilde\gamma,}
and hence
\eqn\solruniv{\eqalign{{\cal G}(Q^2,N)&={1\over 2\pi i}
\int_{\half -i\infty}^{\half+i\infty}
{1\over \gamma}
\exp(\gamma\ln(Q^2/\Lambda^2)-X_0(\gamma)/(\bar\beta_0 N))
d\gamma
\cr
&\hskip 1in \times\int_{0}^{\infty}
\exp(-\tilde\gamma\ln(Q_0^2/\Lambda^2)+X_0(\tilde \gamma)/
(\bar\beta_0N))d\tilde\gamma
\,g_B(Q_0^2,N)\cr
&={\cal G}_E(Q^2,N){\cal G}_I(Q_0^2,N)g_B(Q_0^2,N).\cr}}

The essential expression, 
$\exp(X_0(\gamma)/(\bar\beta_0N))$, contains singularities at all positive 
integers, and ${\cal G}_I(Q_0^2,N)$ is
not properly defined, since the integrand has singularities 
lying along the line of integration. These are due to the divergence of the 
coupling at low $k^2$ and can only be removed by some 
infrared regularization. However, since this factor is independent of 
$Q^2$, it does not contribute at all to the evolution of the
structure function, and is irrelevant for the evolution. The 
function ${\cal G}_E(Q^2,N)$ is determined by the singularities of 
$\exp(-X_0(\gamma)/(\bar\beta_0 N))$ in the $\gamma$ plane.
This leads to a fundamental difference between the cases of the fixed and 
running couplings. Whereas previously the leading singularity was a pole 
at $(\bar\alpha_s/N)\chi(\gamma)=1$, i.e. at $\gamma\to \half$ as $N \to 
4\ln 2\bar\alpha_s$, now the leading singularity is an cut at
$\gamma=0$: there is no powerlike behaviour in $Q^2$. Similarly, the 
branch point in the $N$ plane at $4\ln 2\bar\alpha_s$ has become an essential
singularity at $N=0$: there is no powerlike behaviour in $x$. The introduction
of the running of the coupling has changed the character of the
solution completely. 

Acknowledging that the only real information contained in ${\cal G}_E(N,Q^2)$
is on the evolution of the structure function, i.e. defining 
\eqn\evol{{d \ln{\cal G}(N,Q^2) \over d\ln(Q^2)} =
{d\ln {\cal G}_E(N,Q^2) \over d \ln(Q^2)}\equiv \Gamma(N,Q^2).}
${\cal G}_E(N,Q^2)$ gives us an entirely perturbative effective
anomalous dimension governing the evolution of the gluon structure function. 
The time-honoured technique for solving for ${\cal G}_E(N,Q^2)$ 
is to expand the
integrand in \solruniv, about the saddlepoint. This results in a
contour of integration parallel to the imaginary axis, with real part
$\to \half$ for the small $x$ solutions. Using this results in
an anomalous dimension

\eqn\nloevol{\Gamma(N,Q^2)=\gamma_0(\bar\alpha_s(Q^2)/N)+\sum_{n=1}^{\infty}
(-\beta_0\alpha_s(Q^2))^n
\tilde\gamma_n(\bar\alpha_s(Q^2)/N),}
i.e., the effective anomalous dimension is the naive leading order result
with coupling at scale $Q^2$ plus a series of corrections in increasing powers
of $-\beta_0\alpha_s(Q^2)$. However, each of the $\tilde
\gamma(\bar\alpha_s(Q^2)/N)$ is singular at $N=\lambda(Q^2)$, and the
power of the singularity increases with increasing $n$
\ref\xscale{R.S. Thorne, 
\PR \vyp{D60}{1999}{054031}.}. Hence,
although the series for the resulting splitting function is in the
small quantity $\alpha_s(Q^2)\beta_0$, the accompanying coefficients are
progressively more singular as $x\to 0$. The saddlepoint approximation
is therefore not a reliable result as $x\to 0$ and explicit
investigation reveals that it is only really quantitatively useful when
$\bar\alpha_s(Q^2)\ln(1/x)$ is so small that the effective anomalous
dimension is effectively the LO in $\alpha_s$ part
$xP(x)=\bar\alpha_s(Q^2)$. Therefore any calculations of
the anomalous dimension which rely on an expansion about the
saddle-point lead to very inaccurate and misleading results for small $x$. 

This instability is not surprising if one examines the integrand along
the saddle-point contour of integration, noting that it is very
different from the Gaussian form the saddle-point method assumes \xscale,  
and also if one notes
that it is an expansion obtained from approaching $\gamma=\half$
and in terms of functions of $N$ which are singular at
$N=\lambda(Q^2)$, whereas we know that the full solution no longer
sees these points as anything special. In fact, the knowledge of
the singularity structure of the integrand implies that $\gamma=0$ is
a more fruitful point on which to concentrate. Prompted by this we may
move the contour of integration to the left and simultaneously use the
property that the integrand dies away very quickly at infinity to
close the contour so that it simply encloses the real axis for
$\gamma<0$. It is then useful to express $\chi_0(\gamma)$ in the form  
\eqn\expan{\chi_0(\gamma)=1/\gamma+\sum_{n=1}^{\infty}2\zeta(2n+1)\gamma^{2n},}
which is strictly valid only for $|\gamma|<1$. Doing this we may write
\eqn\expX{X_0(\gamma)= \ln(\gamma)+\gamma_E+
\sum_{n=1}^{\infty}2{\zeta(2n+1)\over
2n+1}\gamma^{2n+1},}
and the integrand for ${\cal G}_E(N,Q^2)$ becomes
\eqn\expint{\gamma^{-1/(\bar\beta_0N)-1}\exp\biggl(\gamma t
-{1\over(\bar\beta_0N)}(\gamma_E+\sum_{n=1}^{\infty}a_n\gamma^{2n+1})\biggr),}
where $t=\ln(Q^2/\Lambda^2)$ and $a_n=2\zeta(2n+1)/(2n+1)$. The
contribution to the integral from $0 \to -\infty +i\epsilon$ is now
the same as
that from $-\infty-i\epsilon \to 0$ up to a phase factor, and we may write
\eqn\contour{{\cal G}_E(N,t)=-\sin\biggl({\pi\over(\bar\beta_0N)}\biggr)
\exp\biggl(-{\gamma_E\over(\bar\beta_0N)}\biggr)
\int_{-\infty}^0 
\gamma^{-1/(\bar\beta_0N)-1}\exp\biggl(\gamma t
-{1\over (\bar\beta_0N)}\sum_{n=1}^{\infty}a_n\gamma^{2n+1}\biggr)d \gamma,}
where the integral has to be understood as an analytic continuation,
since there are singularities along the real axis, and strictly
speaking the integrand is well defined only for $\gamma > -1$. Since
the factor of $\exp(\gamma t)$, is present this latter point leads to an
ambiguity of order $\exp(-t)$, i.e. ${\cal O}(\Lambda^2/Q^2)$ into the
value of ${\cal G}_E(N,t)$. 

In order to evaluate the above integral it is convenient to let
$y=\gamma t$, resulting in 
\eqn\contour{\eqalign{{\cal G}_E(N,t)=
-\sin\biggl({\pi\over(\bar\beta_0N)}\biggr)
&\exp\biggl(-{\gamma_E\over(\bar\beta_0N)}\biggr)\cr
&\hskip -0.8in t^{1/(\bar\beta_0N)}
\int_{-\infty}^0 
y^{-1/(\bar\beta_0N)-1}\exp(y)\exp\biggl(
-{1\over(\bar\beta_0N)}\sum_{n=1}^{\infty}a_n(y/t)^{2n+1}\biggr)d \gamma.\cr}}
The latter exponential may be expanded as a power series in $y/t$ and
the integral evaluated using the standard result that  
\eqn\gamdef{(-1)^n\Gamma(-1/(\bar\beta_0N)+n)=\int_{-\infty}^0 
y^{-1/(\bar\beta_0N)-1}\exp(y)y^{n}d \gamma,}
and hence  
\eqn\result{{\cal G}_E(N,t)=-\sin\biggl({\pi\over(\bar\beta_0N)}\biggr)
\exp\biggl(-{\gamma_E\over(\bar\beta_0N)}\biggr)
t^{1/(\bar\beta_0N)}
\sum_{n=3}^{\infty}A_n(1/(\bar\beta_0N))t^{-n}(-1)^n
\Gamma(-1/(\bar\beta_0N)+n).}
This result was noted in \ref\alanjan{A.D. Martin and J. Kwiecinski,
\PL \vyp{B353}{1995}{123}.}, as was the fact
that it may be simplified by using the relationship that as $N\to 0$, 
$\Gamma(-1/(\bar\beta_0N)+n)\to \Gamma(-1/(\bar\beta_0N))
(-1/(\bar\beta_0N))^{n}$. However, it is important to notice the more
general result that for all $N$ 
\eqn\deltadef{(-1)^n\Gamma(-1/(\bar\beta_0N)+n)= \Gamma(-1/(\bar\beta_0N))
\Delta_n(-1/(\bar\beta_0N)),}
where 
\eqn\deltadefi{\Delta_n(-1/(\bar\beta_0N))=(-1)^n\sum_{m=1}^n (-1)^md_{nm} 
(\bar\beta_0N)^{-m},}
where $d_{nm}$ are positive coefficients and $d_{nn}=1$. Hence,
ignoring the common factor of
$-\sin(\pi/(\bar\beta_0N))\Gamma(-1/(\bar\beta_0N))
\exp(-\gamma_E/(\bar\beta_0N))$, which has no $t$
dependence, and is irrelevant for the anomalous dimension, 
\eqn\resulti{{\cal G}_E(N,t)=t^{1/(\bar\beta_0N)}\biggl(1+
\sum_{n=3}^{\infty}A_n(1/(\bar\beta_0N))t^{-n}\Delta_n(-1/(\bar\beta_0N))
\biggr)}
where the $A_n$ are simply calculable from the expansion of $\exp\biggl(
-1/(\bar\beta_0N)\sum_{n=1}^{\infty}a_n(y/t)^{2n+1}\biggr)$. 
The common factor of
$t^{1/(\bar\beta_0N)}$ is the well-known double-leading-log result
coming from just the LO $\alpha_s(Q^2)/N$ part of the anomalous
dimension. Multiplying this we have an expansion as a power series in $1/t$ or
equivalently in $\alpha_s(Q^2)$. In fact 
\eqn\expanbit{t^{-n}\Delta_n(-1/(\bar\beta_0N))=(\bar\alpha_s(Q^2)/N)^n
\sum_{m=1}^nd_{nm}(-\beta_0\alpha_s(Q^2))^{n-m}(\bar\alpha_s(Q^2)/N)^{m-n}.}
This explicitly
demonstrates that we obtain a set of running coupling corrections to a
LO result. Substituting this expression for
${\cal G}_E(N,t)$ in \evol\ one obtains an expression for the anomalous
dimension as a power series in $\alpha_s(Q^2)$, where at each order we
have the leading divergence in $1/N$ plus a sum of running
coupling correction type terms.
With a little work one may regain the whole leading
$\gamma_0(\alpha_s(Q^2)/N)$ (though it is necessary to keep some
subleading terms in the $\Delta_n$ to do this), along with a tower of
terms which are subleading in powers of $\beta_0\alpha_s(Q^2)$ to this
leading anomalous dimension, i.e. one obtains all the corrections to
this naive LO anomalous dimension due to the running of the coupling.   

The general features of this full, running coupling BFKL anomalous
dimension may be appreciated quite easily.
The important fact to note is that although the
$\Delta_n(-1/(\bar\beta_0N)))
\to (1/(\bar\beta_0N))^n$ as $N\to 0$, the function oscillates with
$1/(\bar\beta_0N)$, and remains very much smaller in magnitude than this
asymptotic form until very small $N$, roughly until $1/N>n$. This
coupled with the accompanying factor of $t^{-n}$ means that for
reasonable $t$, i.e $t>5$ ($Q^2\gsim 1\Gev^2$), only the first $5$ or
so terms in \resulti\ contribute for $N>0.25$. Hence, to a very good
approximation    
\eqn\resulti{{\cal G}_E(N,t)=t^{1/(\bar\beta_0N)}\biggl(1-
{2\zeta(3)\over 3(\bar\beta_0N)t^3}\Delta_3(-1/(\bar\beta_0N))
-{2\zeta(5)\over 5(\bar\beta_0N)t^5}\Delta_5(-1/(\bar\beta_0N))\biggr),}
and in fact the smallness of the coefficient makes even the $t^{-5}$
term almost negligible. ${\cal G}_E(N,t)$ initially grows as $N$ falls due to
the $t^{1/(\bar\beta_0N)}$ term. However, for $N\sim 0.6$ the negative
contribution from the $t^{-3}$ term starts to contribute and ultimately
drives the gluon structure function 
to negative values. The result is shown in
\fig\gluon{The $Q^2$-dependent part of the gluon structure function,
${\cal G}_E(N,t)$, and $d{\cal G}_E(N,t)/dt$ as a function of $N$ for
$t=6$ ($Q^2 \sim 8\Gev^2$). The $Q^2$-independent factor of
$-\sin(\pi/(\bar\beta_0N))\Gamma(-1/(\bar\beta_0N))
\exp(-\gamma_E/(\bar\beta_0N))$ is included in both in order to 
produce a more
smoother $N$-dependent normalization of the functions.}. 
$d{\cal G}_E(N,t)/dt$ may simply be
evaluated also using \resulti, and shows the same general shape, but
does not become negative until a slightly lower value of $N$ as also
seen in \gluon.  
Hence the anomalous dimension develops a pole at a finite value of
$N$, given quite accurately by 
\eqn\pole{t^3=
{2\zeta(3)\over
3(\bar\beta_0N)}\biggl({1\over(\bar\beta_0N)^3}-{3\over(\bar\beta_0N)^2} 
+{2\over(\bar\beta_0N)}\biggr),}
where we use the explicit form of
$\Delta_3(-1/(\bar\beta_0N))$. The value of $N$ for this leading pole
is shown as a function of $t$ in \fig\intercepts{The positions of the
leading poles in the anomalous dimensions for the gluon structure
function at LO, for $F_L$ at LO and for $F_L$ at NLO.},
and for the sort of values relevant at HERA is $\sim 0.25$. 
Going to $N<0.25$ higher order terms become
important, and the positive $1/((\bar\beta_0N)^2t^6)
\Delta_6(-1/(\bar\beta_0N))$ term pulls ${\cal G}_E(N,t)$ back to positive
values, and another pole, with opposite sign residue, appears in
$\Gamma(N,t)$. At even lower $N$ the analytic expression eventually
breaks down, but numerical results show a series of poles becoming
closer together. Nevertheless, the position of the leading pole is
essentially determined by the first handful of terms in the
power series in $\alpha_s(Q^2)$ for ${\cal G}_E(N,t)$, and hence so is the
asymptotic behaviour of the small $x$ splitting function, i.e. 
$\sim x^{-0.25}$. Hence, the
introduction of the running coupling has a dramatic effect on the
singularity structure of the LO BFKL anomalous dimension, turning the
cut into a series of poles, and changing the
position of the rightmost singularity by a factor of $\sim 0.4$. This
result of the pole in the anomalous dimension has been already noted
in \ref\ciafsalam{M. Ciafaloni, D. Colferai and G.P. Salam,
\PR\vyp{D60}{1999}{114036}.} using
numerical techniques, and in the context of a resummed NLO anomalous
dimension. Here I particularly stress the huge modification of the
naive LO BFKL anomalous dimension. This huge change is apparent over a wide
range of $N$, and in
\fig\anom{a. The anomalous dimensions for the gluon structure
function at LO and for $F_L$ at LO plotted as functions of $N$ for
$t=6$ ($Q^2 \sim 8\Gev^2$). Also shown is the ${\cal
O}(\alpha_s(Q^2))$ contribution $\bar\alpha_s(Q^2)/N$.
b. The anomalous dimensions for for $F_L$
at LO and at NLO plotted as functions of $N$ for
$t=6$.}.a I show the 
anomalous dimension
as a function of $N$ for all values right of the leading
singularity. As one sees, it is rather closer to the simple
$\alpha_s(Q^2)/N$ expression than to the naive BFKL result. 
 
Here I should comment on the limit of the analytic expression. As
noted, it involves a series expansion not valid over the whole contour
of integration. This is reflected in the fact that the overall
magnitude of the $\Delta_n(-1/(\bar\beta_0N))$ increases like $n!$ in
general. This means that the series in \resulti\ is actually
asymptotic, although it is an oscillating series, so in principle is
unambiguously resummable (i.e. the integral in \solruniv\ does exist).
However, the greatest accuracy may be obtained from \resulti\ by
truncating the series at order $n_0\sim t$. In practice I always use
$n_0=5$. Substituting the truncated expression for ${\cal G}_E(N,t)$ then
results in an infinite series in $\alpha_s(Q^2)$ for $\Gamma(N,t)$
which is convergent for any given $N$ right of the leading pole. 
The accuracy of the analytic
expression can be found by comparing with results obtained from
evaluating \solruniv\ using numerical integration, and for the gluon
structure function for $N$ to the right of the leading pole is found to
be rather better than $0.1\%$ for $t=6$ and falls like
$\exp(-t)$. 

In order to investigate the quantitative effect of the BFKL anomalous
dimension on structure function evolution it is necessary to calculate
the BFKL splitting
function as a function of $x$. This is where an analytic expression for
the anomalous dimension is particularly useful. A series of
numerically obtained values of $\Gamma(N,t)$ allows an approximate
determination of $P(x,t)$, but it is very difficult to be accurate,
especially for the wildly oscillating functions of $1/N$ which do in
fact make up ${\cal G}_E(N,t)$. However, I now
have an explicit series for $\Gamma(N,t)$ in powers of
$\alpha_s(Q^2)$, obtained from the truncated expression for 
${\cal G}_E(n,t)$. 
The $N$-dependent functions at each power of 
$\alpha_s(Q^2)$ of course
become larger at small $N$ as the series progresses, and to reach small
enough $x$ more and more terms are needed. However, at a fixed value of
$N$ there is no such growth, and the same is therefore true for fixed
$x$. Hence,
one only needs to work to a finite order. Limiting oneself to
$x>10^{-5}$ and $t>4.5$, the suppression of the
$\Delta_n(-1/(\bar\beta_0N))$
is quite significant and seventh order in $\alpha_s(Q^2)$ is easily
sufficient. The splitting function for $t=6$ is shown in 
\fig\splitf{a. The splitting functions $xP_{gg}(x)$ and $xP_{LL}(x)$
plotted as a function of $x$ for
$t=6$ ($Q^2 \sim 8\Gev^2$). Also shown is the ${\cal O}(\alpha_s(Q^2))$ 
contribution $\bar\alpha_s(Q^2)$, and the naive LO BFKL splitting
function with coupling $\alpha_s(Q^2)$.
b. The splitting functions $xP^{LO}_{LL}(x)$ and $xP^{NLO}_{LL}(x)$ 
plotted as a function of $x$ for
$t=6$.}.a. One
sees that it is hugely suppressed compared with the naive LO BFKL
splitting function, and is even lower than the ${\cal O}(\alpha_s(Q^2)$)
contribution for $x$ between about $0.1$ and $0.001$.
The fact that there is deviation from the standard NLO
in $\alpha_s(Q^2)$ splitting functions tells us that BFKL
influenced structure functions are important. 
The ultraviolet renormalon
contribution is approximated by constructing an $x$-space function
which matches the $N$-space results for a variety of $N$. 
Although this contribution
turns out to be a larger fraction of the total than in $N$-space,
it still only makes a very small correction to the evolution. This
will be discussed more in a subsequent paper \ref\future{R.~S.~Thorne,
in preparation.}.      

\newsec{Small $x$ Structure Functions.}

It is important to realize that all of the above results are in a
sense ambiguous because they deal with a
particular way of defining the gluon parton distribution. It is
defined in the natural way for a discussion of BFKL physics, but
nonetheless is really an intrinsically factorization-scheme-dependent
quantity. One may define a real structure function by including a hard
scattering cross section at the top of the gluon ladder. This modifies
\gluondef\ to 
\eqn\strcfun{{\cal F}_{i}(Q^2,N)=\alpha_s\int_{0}^{\infty}{dk^2\over k^2} 
\sigma_{i,g}(k^2/Q^2) f(N,k^2,Q_0^2)
g_B(N,Q_0^2).}
Pulling out the overall factor of $\alpha_s$ 
and taking the Mellin transformation of \strcfun\ with respect to 
$(Q^2/\Lambda^2)$ leads to the simple expression
\eqn\strcfunmell{\tilde {\cal F}_{i}(\gamma,N)= 
h_{i,g}(\gamma) \tilde {\cal G}(\gamma,N).}
Thus we may solve for ${\cal F}_{i}(N,t)$ in the same way as for 
${\cal G}(N,t)$ obtaining the same divergent $Q^2$-independent part and
a $Q^2$-dependent part given by solving 
\eqn\solrunf{{\cal F}_{E,i}(N,t)={1\over 2\pi i}
\int_{\half -i\infty}^{\half+i\infty}
{h_{i,g}(\gamma)\over \gamma}
\exp(\gamma t-X_0(\gamma)/(\bar\beta_0 N))
d\gamma.}
We may proceed as with the gluon structure function
by expanding the $h_{i,g}(\gamma)$ (which were calculated in
\ref\cathaut{S. Catani and F. Hautmann, \PL \vyp{B315}{1993}{157}; \NP
\vyp{B427}{1994}{475}.}) as a power series about
$\gamma=0$. This results in an expression
\eqn\resultif{{\cal F}_{E,i}(N,t)=t^{1/(\bar\beta_0N)}\biggl(1+
\sum_{n=1}^{\infty}B_{i,n}(1/(\bar\beta_0N))t^{-n}\Delta_n(-1/(\bar\beta_0N))
\biggr),}
where the $B_{i,n}(1/(\bar\beta_0N))$ are now determined not only by
the power series in $\gamma$ obtained from the expansion 
of $X_0(\gamma)$, but also from the expansion of
$h_{i,g}(\gamma)$. In particular they now contain parts at zeroth
order in $1/(\bar\beta_0N)$. In order to draw the most direct analogy to the
gluon we define 
\eqn\physlong{\Gamma_{LL}(N,t)={d \ln({\cal F}_L(N,t))\over d t}.}  
All the terms in the expression
for this physical anomalous dimension,
determined from \resultif, are part of the standard LO
$\Gamma^0_{LL}(\alpha_s(Q^2)/N)$ or are subleading by powers of
$\beta_0\alpha_s(Q^2)$ to this, i.e. again we obtain the naive LO
result plus running coupling induced corrections to this, and the
``coefficient function'' $h_{i,g}(\gamma)$ contributes only to the
running coupling corrections.  

$\Gamma_{LL}(N,t)$ is shown along with the gluon anomalous dimension in 
\anom.a. Clearly the effect of the additional coefficient function, and
hence additional running coupling corrections, is to make
$\Gamma_{LL}(N,t)$ dip significantly below the simple $\bar\alpha_s(Q^2)/N$
and to reduce the value of the intercept compared to the
gluon structure function. This is reflected in the effective
splitting function $P_{LL}(x,t)$ which is shown in \splitf.a. This time
the dip below the ${\cal O}(\alpha_s(Q^2))$ part is far more
pronounced. Also, going to low enough $x$, we see that the splitting
function turns over again, showing that the subleading poles in the
anomalous dimension may have large residues compared to the leading
pole, and the increase in $P_{LL}(x)$ with decreasing $x$ is not
monotonic. This illustrates that as far as phenomenology at HERA, or
any forseeable collider, is concerned, the value of the intercept for
the anomalous dimension is simply not relevant to the evolution of
structure functions.
I should note that in the case of the physical anomalous
dimension and splitting function the power corrections due to the
nonconvergence of the series are somewhat larger, about $1\%$ in
$N$-space for $t=6$, and lead to a small, but not insignificant effect
on the evolution in $x$-space, and must be accounted for. This will be
discussed more in \future.   

One can follow exactly the same procedure for the other physical
anomalous dimension
\eqn\physanomtwo{{\partial {\cal F}_2(N,Q^2)\over \partial \ln Q^2} = 
\Gamma_{2L}(Q^2,N) {\cal F}_L(N,Q^2),}
obtaining qualitatively very similar results. Again the splitting
function initially dips as $x$ decreases, but grows again in the same
way as $P_{LL}(x,t)$ at lower $x$. This has extremely important
implications for the evolution of $F_2(x,Q^2)$ in the HERA range, and
this will be presented in detail in \future.

\newsec{NLO Corrections.}   

So far I have demonstrated that using $\alpha_s(k^2)$ in the BFKL
equation, as in \bfklruni, has a profound effect on the form of the
solution for the anomalous dimension. However, it is necessary to
check that the results presented are not severely modified by the
inclusion of the NLO kernel, i.e. the perturbative calculations are
stable, and also to justify that the choice of scale in the coupling
is correct, or at least leads to accurate results. The NLO kernel was
presented in \NLOBFKLlf\ and the way in which to solve at NLO with a running
coupling was presented in \ref\ciafpcol{M. Ciafaloni 
and D. Colferai, \PL \vyp{452}{1999}{372}.}. 
Writing the NLO equation as 
\eqn\NLOBFKLalti{f(k^2,Q_0^2)= f_I(k^2,Q_0^2)
+\biggl({\bar\alpha_s(k^2) \over N}\biggr)
\int_0^{\infty}{dq^2\over q^2}(K_0(q^2,k^2)
-\alpha_s(k^2)K_1(q^2,k^2))f(q^2),}
using just the one-loop coupling leads to a 2nd order differential equation in
$\gamma$-space
\eqn\bfklrunnlotot{{d^2 \tilde f(\gamma,N)\over d \gamma^2}={d^2\tilde 
f_I(\gamma, Q_0^2) \over
d\gamma^2}-{1\over \bar\beta_0 N} {d (\chi_0(\gamma)\tilde 
f(\gamma,N))\over d\gamma}-{\pi\over 3\bar\beta^2_0 N}\chi_1(\gamma)
\tilde f(\gamma, N).}
This can be solved in a very similar way to LO,
i.e. it factorizes into the same form as \solruniv\
with $Q^2$-dependent part given by
\eqn\solrunnloii{{\cal G}_{E,NLO}(N,t)={1\over 2\pi i}
\int_{\half -i\infty}^{\half+i\infty}
{1\over \gamma}
\exp(\gamma t-X_{NLO}(\gamma,N)/(\bar\beta_0 N))
d\gamma.}
However, $X_{NLO}(\gamma,N)$ is rather more
complicated than the previous $X_0(\gamma)$.
It can be expressed in the form 
\eqn\solrunnloiii{X_{NLO}(\gamma,N)=\int_{\half}^{\gamma}
\chi_{NLO}(\hat\gamma,N)d\hat\gamma,}
where $\chi_{NLO}(\gamma,N)$ can be written as a power series in $N$ 
beginning at zeroth order with $\chi_0(\gamma)$. As seen in \ciafpcol,
though here ignoring any resummations in $N$, the explicit form is 
\eqn\solrunnloiv{\chi_{NLO}(\gamma,N) =
\chi_0(\gamma)-N{\chi_1(\gamma) \over \chi_0(\gamma)} +{N^2 \over
\chi_0}\biggl(-\biggl({\chi_1(\gamma)\over
\chi_0(\gamma)}\biggr)^2-\beta_0 \biggl({\chi_1(\gamma)\over 
\chi_0(\gamma)}\biggr)'\biggr) +\cdots,}
where $\chi_2(\gamma)$ would also appear at order $N^2$, had I included
it. I shall generally ignore all but the first two terms. 

Firstly, I shall address the choice of scale. It was known in 
\ref\ciafrun{G. Camici and M. Ciafaloni, \PL\vyp{B386}{1996}{341}.}
that the correct scale was really $(k-q)^2$, but that $k^2$ could
be used, leading to a part of the NLO kernel which is proportional to
$\beta_0$, i.e. there is a contribution to $\chi_1(\gamma)$ of the
form $\half\bar\beta_0(\chi_0^2(\gamma) +\chi'_0(\gamma))$. Substituting
this into \solrunnloii\ leads to a contribution in the integrand of 
the form $\exp(\half(\ln(\chi_0(\gamma))+X_0(\gamma)))$. 
This can be expressed as a power
series which at low orders is $1+1.6\gamma^3+1.24\gamma^5$. Hence, this
scale-induced factor has the same form as the $h_{i,g}(\gamma)$, and not
surprisingly results in additional running coupling corrections to the
anomalous dimensions. However, the terms in the series do not start
until third order, have small coefficients, and have an effect much
smaller than $h_{i,L(2)}(\gamma)$. Hence, the correction for this
``incorrect'' choice of scale is very small, though in principle it seems as
though the factor just considered should really be taken as part of the LO
result, since it just gives running coupling corrections only.\foot{In
fact, the anomalous dimensions
and splitting functions already presented in this paper 
contain corrections from
this factor, though this makes only a very small difference.}

So the choice of $\alpha_s(k^2)$ is in practice very reliable, and
may easily be corrected for. We must now consider the rest of the NLO
correction to the kernel, which is much larger. Here we have an
ambiguity in precisely what the NLO calculation means. Do we simply
solve \NLOBFKLalti, producing the infinite series in \solrunnloiv? 
Do we truncate $\chi_{NLO}(\gamma,N)$ after the second term, and if
so do we use the whole of $\exp(1/\bar\beta_0\int_{\half}^{\gamma}
(\chi_1(\hat\gamma)/\chi_0(\hat\gamma))d\hat\gamma)$ or just expand it
out to first order? There are particular problems associated with all
choices. I choose the NLO definition such that the anomalous
dimension receives only corrections which are one power of
$\alpha_s(Q^2)$ down on the leading order one, i.e. the LO
anomalous dimension is of the form $\Gamma^0(\bar\alpha_s(Q^2)/N,
\beta_0\alpha_s(Q^2))$ and the NLO corrected one is of the form 
$\Gamma^0(\bar\alpha_s(Q^2)/N,\beta_0\alpha_s(Q^2))+\alpha_s(Q^2)
\Gamma^1(\bar\alpha_s(Q^2)/N,\beta_0\alpha_s(Q^2))$. Roughly speaking,
this involves keeping only the first two terms in \solrunnloiv, and
expanding $\exp(1/\bar\beta_0\int_{\half}^{\gamma}
(\chi_1(\hat\gamma)/\chi_0(\hat\gamma))d\hat\gamma)$, out to just first
order in $1/\bar\beta_0$. However, the 
part of $(\chi_1(\hat\gamma)/\chi_0(\hat\gamma))$ 
behaving like $1/\gamma$ must be treated very carefully. Details will
be presented in \future. 

Solving for the NLO anomalous dimension using the same techniques 
as at LO results in the NLO correction. Unlike the case of fixed
coupling, or the naive results of the saddle-point evaluation, these
corrections are rather small. The positions of the leading poles in
the anomalous dimensions are shown in \intercepts, and one can see
that they change from about $0.21$ for $\Gamma_{LL}$ at LO to $0.16$
at NLO, and that the $Q^2$-dependence reduces a little. 
Similarly the anomalous dimension $\Gamma_{LL}(N,t)$ over a wide range
of $N$ shows
only a very small change going from LO to NLO. (I use the physical
anomalous dimensions to avoid any ambiguity. The results are very
similar for the gluon distribution.)
This is shown in \anom.b where the
part of the NLO anomalous dimension at first order in $\alpha_s(Q^2)$,
i.e $-0.935\alpha_s(Q^2)$,
is not included, since this should properly be included at LO in a
combined leading order in $\alpha_s(Q^2)$ and $\alpha_s(Q^2)\ln(1/x)$
expansion scheme. Alternative definitions of NLO lead to very similar
results except at very high values of $N$. Note that the NLO
correction is actually positive for $N\sim 0.9$ - very different from
the case where running coupling corrections are not included and all
NLO corrections are negative. I should also note that the ultraviolet
renormalon contribution is a far larger proportion of the NLO correction
than it is of the LO contribution, being typically $10\%$ for
$t=6$, and needs to be accounted for \future.           
 
One can also make the transformation to $x$-space and calculate the
NLO corrected splitting function. This is shown for $t=6$ in
\splitf.b, where the 
contributions $\propto \delta(1-x)$ both from the
${\cal O}(\alpha_s(Q^2))$ part and the running coupling corrections to this
are absent. The latter of these is a very small contribution. The NLO
corrected splitting function is not too different from that at LO, as
one can see. However, the real import of the NLO corrections as far as
physics is concerned is the effect it has on the evolution of the
structure function. This is demonstrated in \fig\evolution{The values
of $d F_L(x,Q^2)/d\ln Q^2$, for $F_L(x,Q^2)=x^{-0.2}(1-x)^6$, due to the 
LO splitting functions $P^{LO}_{LL}(x)$ and the NLO splitting
function $P^{NLO}_{LL}(x)$, 
plotted as a function of $x$ for
$t=6$ ($Q^2 \sim 8\Gev^2$). Also shown is the evolution due to the
${\cal O}(\alpha_s(Q^2))$ contribution $P(x)=\bar\alpha_s(Q^2)/x$.} 
where the evolution of a suitable model
for the structure function $F_L(x,Q^2)$, i.e. $(1-x)^6x^{-0.2}$, is
shown both for the LO running coupling splitting function, and for the
NLO corrected one (all $\delta(1-x)$ contributions other that at
first order in $\alpha_s(Q^2)$ one are included). As one sees, at this
(rather low) value of $t$, i.e. $Q^2\sim 8\Gev^2$, the effect of the
NLO corrections is only of order $10\%$, and is positive for $x\sim
0.01$. Also shown for comparison is the contribution from the
$\bar\alpha_s(Q^2)/x$ term alone. The running coupling
BFKL splitting function leads to slightly quicker evolution than this
latter contribution for $x\gsim 0.01$, but for $x\sim 0.0001$ the
evolution is significantly suppressed.       
        
Hence, the NLO corrections to the running coupling BFKL derived
splitting function are well under control, both in terms of the
asymptotic powerlike behaviour of the splitting functions and in terms
of the evolution in the range currently accessible to experiments.
Beyond the running coupling corrections no further resummations are
necessary, or even useful. This is very much in contrast to the case
where both ends of the gluon ladder are associated with a hard scale,
the so-called ``single scale'' processes. In this case, as shown in    
\ref\gavin{G.P. Salam, JHEP \vyp{9807}{1998}{19}.} and developed in
\ciafpcol\ and \ciafsalam, the conventional
BFKL expansion is fundamentally flawed due to high order poles at
$\gamma=0$ and $1$, which need to be resummed. Without resummation,
all calculations are badly behaved over the full range of $N$, not
just at low $N$. In the case of deep inelastic 
scattering the collinear factorization procedure automatically orders
the poles at $\gamma=0$ correctly, and the above problem shows up in
high order poles at $\gamma=1$ only. The anomalous dimension is
totally dominated by the region very close to $\gamma=0$, as this paper
shows, and is very insensitive to effects at $\gamma=1$. Including the
type of resummation in \gavin\ciafpcol\ alters results from the NLO
corrected case by a very small amount, and is possibly no more
influential than the remaining NNLO effects for which it does not account. 
The corrected treatment at $\gamma=1$ is essential if one is attempting to
obtain information about the input form of the gluon,
i.e. ${\cal G}_I(Q_0^2,N)$, but this, along with the whole subject of
single-scale processes, is also plagued by the infrared divergence problem.  
A discussion of such issues can be found in 
\ciafsalam\ref\ciafcolsalam{M.Ciafaloni, D. Colferai and G.P. Salam,
JHEP \vyp{9910}{1999}{017}.}.

\newsec{Conclusions.}

I have shown that an analytic expression for the anomalous dimensions
and splitting functions obtained from the running coupling BFKL
equation may be obtained by a power series solution for the 
$Q^2$-dependent part of the gluon
structure function in terms of $\alpha_s(Q^2)$. This has extremely
good accuracy, with only very small errors, which may be interpreted
as ultraviolet renormalon contributions, and may be calculated
numerically. 
Moreover, I find the remarkable result that the form of
this anomalous dimension and splitting function is almost completely 
determined by only the first
handful ($\sim 5$) of terms in the expansion for the gluon. This is in
complete contrast with the case of fixed coupling, where an all orders
summation is needed, and it would be
interesting to understand the origin of this phenomena.   

My results also prove that the effect of the running of the coupling is to
weaken the asymptotic powerlike growth of the splitting
functions severely compared to the naive results, and even to lower the
splitting function below the $\alpha_s(Q^2)/x$ contribution for
$0.001\gsim x \gsim 0.2$. It also makes the NLO correction to the
splitting functions relatively small, both for the value of the
intercept and for the evolution of structure functions for
$x>10^{-5}$, and therefore stabilizes the perturbative series. 
I also note that a previous conjecture that the effect of the running
coupling in the BFKL equation could be accounted for using an
$x$-dependent scale for the coupling \xscale, resulting in falling coupling
for decreasing $x$, turns out to be largely correct so long as 
the change in the scale of the coupling is moderate compared to the 
scale itself, though it fails when this criterion is not satisfied. 
In practice this condition is identical to that specifying that
diffusion in the fixed coupling BFKL equation is not too large and 
therefore that the virtualities sampled in the running coupling
equation are not too far away
from $Q^2$. This results in the requirement that $t^3\gsim 20
\ln(1/x)$ \ref\mueller{A.H. Mueller, \PL 
\vyp{B396}{1997}{251}.}. 
This covers most of the HERA range for $Q^2> 1\Gev^2$, and
means that a phenomenological analysis using the explicit resummation in
this paper leads to very
similar results to the approach in \xscale. In particular, the input 
$F_L(x,Q_0^2)$ is of the same shape
as $F_2(x,Q_0^2)$ for $Q_0^2$ as low as $1\Gev^2$, in contrast to the
standard NLO-in-$\alpha_s(Q^2)$ approach, and the quality of the fit, 
particularly for small $x$ data, is far better than using the
conventional approach.  A more detailed presentation of such a 
phenomenological study will
appear. However, I can say that the inclusion of corrections to the fixed
order in $\alpha_s(Q^2)$ splitting functions from the running coupling
BFKL equation not only leads to a stable perturbative expansion but
also to a clear
improvement in the comparison to data.  

\bigskip

\noindent{\bf Acknowledgements.}
\medskip
I would like to thank A.D. Martin, A.H. Mueller, R.G. Roberts,
D.A. Ross  and W.J. Stirling for useful discussions.

\vfill 
\eject

\vfill\eject\immediate\closeout\ffile{\parindent40pt
\baselineskip14pt\centerline{{\bf Figure Captions}}\nobreak\medskip
\escapechar=` \input figs.tmp\vfill\eject}

\footatend\vfill\supereject\immediate\closeout\rfile\writestoppt
\baselineskip=14pt\centerline{{\bf References}}\bigskip{\frenchspacing%
\parindent=20pt\escapechar=` \input refs.tmp\vfill\eject}\nonfrenchspacing

\end

\ref\bluvogt{J. Bl\"umlein, V. Ravindran, W.L.
van Neerven and  A. Vogt, Proceedings of DIS 98, 
Brussels, April 1998, p. 211, {\tt hep-ph/9806368}.}%
\nref\ballforte{R.D. Ball 
and S. Forte Proceedings of DIS 98, Brussels, April 1998, p. 770,
{\tt hep-ph/9805315}.}%
\nref\ross{D.A. Ross, \PL \vyp{B431}{1998}{161}.}%
\nref\mplusk{Yu. V. Kovchegov 
and A.H. Mueller, \PL \vyp{B439}{1998}{423}.}%
\nref\levin{E.M. Levin, 
Tel Aviv University, Report No. TAUP 2501-98, 
{\tt hep-ph/9806228}.}%
--\ref\armesto{N. Armesto, J. Bartels and M.A. Braun, \PL 
\vyp{B442}{1998}{459}.}

\nref\hancock{R.E. Hancock and 
D.A. Ross, \NP \vyp{B383}{1992}{575}; \NP \vyp{B394}{1993}{200}.}%
\nref\nik{N.N. Nikolaev and B.G. 
Zakharov, \PL \vyp{B327}{1994}{157}.}%
\nref\levrun{E.M. Levin, \NP \vyp{B453}{1995}{303}.}%
\nref\Andersson{B. Andersson, G. Gustafson and 
H. Kharraziha, \PR \vyp{D57}{1998}{5543}.}%
\nref\haakman{L.P.A. Haakman, O.V. 
Kancheli, and J.H. Koch, \PL \vyp{B391}{1997}{157}; \NP 
\vyp{B518}{1998}{275}.}%
-- 
\ref\diff{J.Bartels and
H. Lotter, \PL \vyp{B309}{1993}{400}\semi
J. Bartels, H. Lotter and M. Vogt
\PL \vyp{B373}{1996}{215}.} 

\ref\BLM{S.J. Brodsky, G.P. Lepage and 
P.B. Mackenzie, \PR \vyp{D28}{1983}{228}.}  
\ref\kti{S. Catani, M. Ciafaloni 
and F. Hautmann, \PL  \vyp{B242}{1990}{97}; \NP \vyp{B366}{1991}{135};
\PL\vyp{B307}{1993}{147}.}\ref\ktii{J.C. Collins and R.K. Ellis, \NP 
\vyp{B360}{1991}{3}.}

\eqn\melltranspd{f(N,Q^2)= \int_0^1\,x^{N}{\rm f}(x,Q^2)dx.}

\liprun\diff\ref\harriman{J.R. Forshaw, P.N. Harriman and
P.N. Sutton, \NP \vyp{B416}{1994}{739}.}\haakman\mueller. 

\eqn\nblm{{d {\cal G}(N,Q^2) \over d\ln Q^2} \approx 
\Gamma(N,\bar\alpha_s(s(N)Q^2)) G(N,Q^2),}
\eqn\xblm{{d G(x,Q^2) \over d\ln(Q^2)} \approx \int_x^1
P(z,\bar\alpha_s(s(z),Q^2))G(x/z,Q^2)dz.}
\eqn\nlosplit{(x/\bar\alpha_s(Q^2))P(x,Q^2)=p^0(\bar\alpha_s(Q^2)\xi)
-\beta_0\alpha_s(Q^2)\hat
p^1(\bar\alpha_s(Q^2)\xi)+{\cal O}((\beta_0\alpha_s(Q^2))^2) 
r(\bar\alpha_s(Q^2)
\xi),}
this is the same as 
\eqn\absorb{(x/\bar\alpha_s(Q^2s(\xi
\bar \alpha_s(Q^2))))P(x,Q^2)=p^0(\bar\alpha_s(Q^2s(\xi
\bar \alpha_s(Q^2)))+{\cal O} 
((\beta_0\alpha_s(Q^2))^2) \hat r(\bar\alpha_s(Q^2)\xi),}
As $x \to 0$
\eqn\losplit{p^0(\bar\alpha_s(Q^2)\xi) \to {1\over 
(56\pi\zeta(3))^{\half}} \exp(\lambda(Q^2)\xi)(\bar\alpha_s(Q^2)\xi)^{-3/2},}
So far I have simply assumed that an accurate way to account for the running
of the coupling in the LO BFKL equation is to use \bfklruni. This is an 
assumption which involves the resummation of an infinite number of terms, i.e.
it assumes that at all orders in $\alpha_s(\mu^2)$ the dominant contribution
to the BFKL equation due to the running coupling is 
\eqn\runassump{{\bar \alpha_s \over  N}(-1)^n
(\beta_0\alpha_s(\mu^2)\ln(k^2/\mu^2))^n 
\int_0^{\infty}{dq^2\over q^2}K^0(q^2,k^2)f(q^2),}
Formally the NLO BFKL equation may be
written as   
\eqn\NLOBFKL{\eqalign{f(k^2,Q_0^2,\bar\alpha_s(\mu^2)/N)= f^0(k^2,Q_0^2)
&+\biggl({\bar\alpha_s(\mu^2) \over N}\biggr)
\int_0^{\infty}{dq^2\over q^2}(K^0(q^2,k^2)\cr
&-\beta_0\alpha_s(\mu^2)
\ln(k^2/\mu^2)K^0(q^2,k^2)-\alpha_s(\mu^2)K^1(q^2,k^2))f(q^2),\cr}}
where $K^1(q^2,k^2)$ can be found in \NLOBFKLlf. 
\eqn\NLOBFKLalt{f(k^2,Q_0^2,\mu^2)= f^0(k^2,Q_0^2)
+\biggl({\bar\alpha_s(k^2) \over N}\biggr)
\int_0^{\infty}{dq^2\over q^2}(K^0(q^2,k^2)
-\alpha_s(\mu^2)K^1(q^2,k^2))f(q^2).}
This is identical to \NLOBFKL\ up to NNLO in $\alpha_s(\mu^2)$ and is
a common way for the NLO BFKL equation to be written.
\eqn\bfklrunnlo{{d \tilde f(\gamma,N)\over d \gamma}={d\tilde 
f^0(\gamma, Q_0^2) \over
d\gamma}-{1\over \bar\beta_0 N} (\chi_0(\gamma)-\alpha_s(\mu^2)\chi_1(\gamma))
\tilde f(\gamma, N),}
\eqn\wrongnloan{\alpha_s{\chi_1(\gamma^0(\bar\alpha_s/N))\over 
-\chi'_0(\gamma^0(\bar\alpha_s/N))} \equiv \alpha_s \gamma^1(\bar\alpha_s/N),}
is often called the NLO-BFKL anomalous dimension.

\eqn\nloanom{\Gamma(N,Q^2/\mu^2)=\gamma^0-
\beta_0\alpha_s\biggl({\partial \gamma^0\over
\partial \ln(\alpha_s)}\ln(Q^2/\mu^2)+  
{\partial \gamma^0 \over \partial\ln(\alpha_s)}
\biggl({-\chi''(\gamma^0) \over
2\chi'(\gamma^0)}-{1\over \gamma^0}\biggr)\biggr)-\alpha_s\gamma^1.}
\eqn\nlospli{xP(x,Q^2) =\bar\alpha_s\exp(\lambda\xi)\biggl(
{0.068 \over (\bar\alpha_s\xi)^{3/2}}-\beta_0\alpha_s\biggl(\biggl({0.188 \over
(\bar\alpha_s\xi)^{\half}}\biggr)\ln(Q^2/\mu^2)+0.69\biggr)-
\alpha_s\biggl({1.18 \over
(\bar\alpha_s\xi)^{\half}}\biggr)\biggr).}
\eqn\nloconfan{-\alpha_s(Q^2)\gamma^1(\alpha_s(Q^2)/N)\equiv  
- \alpha_s(Q^2){\chi_1(\gamma^0(\bar\alpha_s(Q^2)/N))
\over -\chi'_0(\gamma^0(\bar\alpha_s(Q^2)/N))}}

\noindent {\bf Table 1.} \hfil\break

The coefficients in the power series $p^i_{LL}(\bar\alpha_s(Q^2)\xi)=
\sum_{0}^{\infty}a_n
(\bar\alpha_s(Q^2)\xi)^{n}/ n!$ for the
various LO and NLO contributions to the physical splitting function 
$P_{LL}(x,Q^2)$. 

\hfil\vtop{{\offinterlineskip
\halign{ \strut\tabskip=0.6pc
\vrule#&  #\hfil&  \vrule#&  \hfil#& \vrule#& \hfil#& \vrule#& \hfil#&
\vrule#& \hfil#& \vrule#\tabskip=0pt\cr
\noalign{\hrule}
& $n$ && $p^0_{LL}$ && $p^{1,tot}_{LL}$ && $p^{1,\beta}_{LL}$ &&
$p^{1,conf}_{LL}$ &\cr
\noalign{\hrule}
& 0 && 1.00 && 0.23 && -2.00 && 1.57 &\cr
& 1 && 0.00 && 4.38 && 4.15 && 1.60 &\cr
& 2 && 0.00 && 15.87 && 11.32 && 8.29 &\cr
& 3 && 2.40 && 13.41 && -16.18 && 24.25 &\cr
& 4 && 0.00 && 86.26 && 76.03 && 35.31 &\cr
& 5 && 2.07 && 252.92 && 167.34 && 140.81 &\cr
& 6 && 17.34 && 323.08 && -81.51 && 377.69 &\cr
& 7 && 2.01 && 1699.65 && 1472.42 && 713.25 &\cr
& 8 && 39.89 && 4338.69 && 2665.07 && 2553.16 &\cr
& 9 && 168.75 && 7592.65 && 1674.16 && 6470.97 &\cr
& 10 && 69.99 && 33409.13 && 28319.16 && 14435.29 &\cr
& 11 && 661.25 && 79427.26 && 47284.56 && 47746.61 &\cr
& 12 && 1945.31 && 173361.43 && 81792.97 && 118560.14 &\cr
& 13 && 1717.68 && 657395.79 && 543255.72 && 293414.46 &\cr
& 14 && 10643.26 && 1527235.16 && 927749.64 && 905642.90 &\cr
& 15 && 25266.78 && 3833618.50  && 23539999.61 && 2256438.84 &\cr
\noalign{\hrule}}}}\hfil

\vfil
\eject

\noindent {\bf Table 2.} \hfil\break

The coefficients in the power series $p^i_{2L}(\bar\alpha_s(Q^2)\xi)=
\sum_{0}^{\infty}a_n
(\bar\alpha_s(Q^2)\xi)^{n}/ n!$ for the
LO and $\beta_0$-dependent NLO contributions to the physical splitting
function $P_{2L}(x,Q^2)$. 

\hfil\vtop{{\offinterlineskip
\halign{ \strut\tabskip=0.6pc
\vrule#&  #\hfil&  \vrule#& \hfil#&
\vrule#& \hfil#& \vrule#\tabskip=0pt\cr
\noalign{\hrule}
& $n$ && $p^0_{2L}$ && $p^{1,\beta}_{2L}$ &\cr
\noalign{\hrule}
& 0 && 2.50 && -4.00 &\cr
& 1 && 1.00 && 9.39 &\cr
& 2 && 1.00 && 36.60 &\cr
& 3 && 7.01 && 6.27 &\cr
& 4 && 5.81 && 239.73 &\cr
& 5 && 13.40 && 687.03 &\cr
& 6 && 58.11 && 771.35 &\cr
& 7 && 64.74 && 5281.50 &\cr
& 8 && 196.83 && 13213.51 &\cr
& 9 && 649.89 && 24043.80 &\cr
& 10 && 930.65 && 111578.92 &\cr
& 11 && 3034.70 && 265509.09 &\cr
& 12 && 8527.87 && 613964.05 &\cr
& 13 && 15046.02 && 2311855.03 &\cr
& 14 && 48434.53 && 5521425.31 &\cr
& 15 && 124600.51 && 14458201.96 &\cr
\noalign{\hrule}}}}\hfil

\vfil
\eject

\noindent Table 3\hfil\break
\noindent Comparison of quality of fits using full leading order (including 
$\ln (1/x)$ terms) renormalization scheme consistent expression, with BLM 
scale setting and the NLO in $\alpha_s(Q^2)$ fit \mrst. The references to the 
data can be found in \mrst.   
\medskip

\hfil\vtop{{\offinterlineskip
\halign{ \strut\tabskip=0.6pc
\vrule#&  #\hfil&  \vrule#&  \hfil#& \vrule#& \hfil#& \vrule#& \hfil#&
\vrule#\tabskip=0pt\cr
\noalign{\hrule}
& Experiment && data &&$\chi^2$&\omit& \omit &\cr
&\omit&& points && LO(x) &\omit& MRST &\cr
\noalign{\hrule}
& H1 $F^{ep}_2$ && 221 && 149 && 164 &\cr
& ZEUS $F^{ep}_2$ && 204 && 246 && 270 &\cr
\noalign{\hrule}
& BCDMS $F^{\mu p}_2$ && 174 && 241 && 249 &\cr
& NMC $F^{\mu p}_2$ && 130 && 118 && 141 &\cr
& NMC $F^{\mu d}_2$ && 130 && 81 && 101 &\cr
& NMC $F^{\mu n}_2/F^{\mu p}_2$ && 163 && 176 && 187 &\cr
& SLAC $F^{\mu p}_2$ && 70 && 87 && 119 &\cr
& E665 $F^{\mu p}_2$ && 53 && 59 && 58 &\cr
& E665 $F^{\mu d}_2$ && 53 && 61 && 61 &\cr
\noalign{\hrule}
& CCFR $F^{\nu N}_2$ && 66 && 57 && 93 &\cr
& CCFR $F^{\nu N}_3$ && 66 && 65 && 68 &\cr
\noalign{\hrule}
& total && 1330 && 1339 && 1511 &\cr
\noalign{\hrule}}}}\hfil
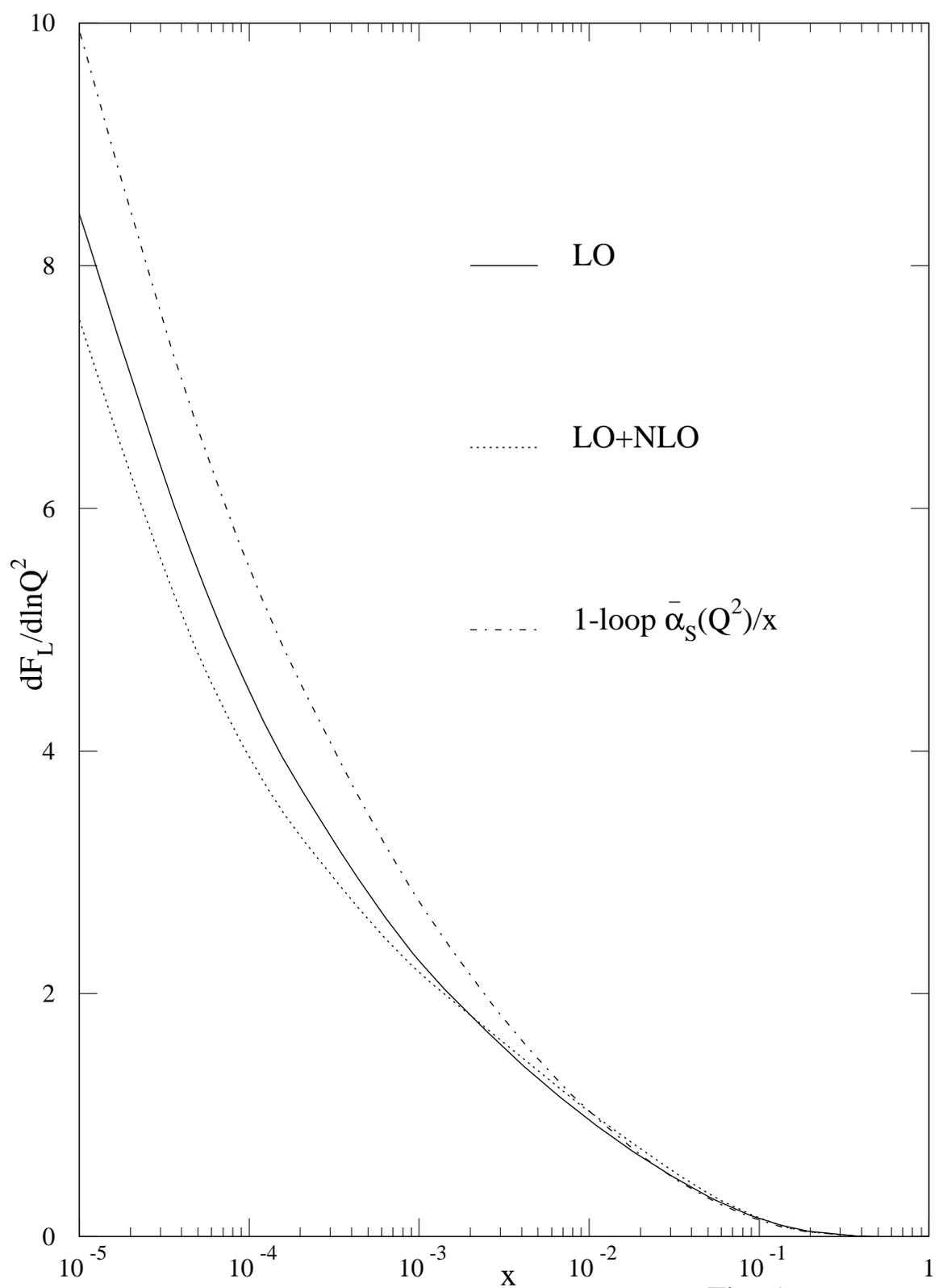
